\documentstyle[prb,aps,epsf,amssymb,floats]{revtex}
\begin{document}
\draft

\title{Band gaps of the primary metallic carbon nanotubes}
\author{Alex Kleiner and Sebastian Eggert}
\address{Institute of Theoretical Physics\\  Chalmers
University of Technology and G\"{o}teborg University\\  
S-412 96 G\"{o}teborg, Sweden \\ }
\date{Submitted: July 6, 2000.  Last Change: November 8, 2000}

\wideabs{
\maketitle
\begin{abstract}
Primary metallic, or small gap semiconducting nanotubes, are tubes
with band gaps that arise solely from breaking the bond symmetry due 
to the curvature. 
We derive an analytic expression for these
gaps by considering how a general symmetry breaking opens a gap
in nanotubes with a well defined chiral wrapping vector.
This approach provides
a straightforward way to include all types of symmetry breaking effects, 
 resulting in a simple unified gap equation as a function of
chirality and deformations.
\end{abstract}

\pacs{PACS numbers: 71.20.Tx, 61.48.+c, 73.61.Wp}}


Recently, individual single wall nanotubes (SWNT) exhibiting small band gaps
of the order of 10 meV where observed for the first time.\cite{zhou 2000}
SWNT's can be classified according to their electronic
band gap into three groups: semiconductors, small
gap semiconductors and real metals. 
A semiconducting gap arises when the graphite
Fermi points are not allowed in the tube's Brillouin zone, which is 
given by distinct quantization lines 
 according to the tube's circumferential boundary conditions.
Such a gap is of the order of 1 eV and 
was predicted to scale with $1/\rm R$ where $\rm R$ is the tube's 
radius,\cite{white 93,jishi 94,mintmire 93} a prediction that
 was later verified
 experimentally by a Scanning Tunneling Microscopy measurement 
of the density of states.\cite{wildoer 98,odom 98,venema 2000}
 The graphite Fermi points lie on a quantization line if 
  ${\rm  mod}(\frac{n-m}{3})=0$,\cite{white 93,hamada 91,saito 92,jishi 93}
where
$n$~and $m$~are the two integers defining the tube's chiral vector.
Tubes that satisfy this condition are called primary metallic.
In a real
nanotube, however, effects of curvature and deformation break the
nearest neighbor bond symmetry, resulting in a shift of the two
distinct Fermi points of graphite which lie at the corners of the
hexagonal first Brillouin zone where the bonding and antibonding bands
are degenerate ($K$-points).
 This shift may open a gap depending on the position of the new 
$K$-points 
relative to the circumferential quantization line. These gaps
 are about the value of room temperature and thus coined small
gap semiconductors. 

Much interest has been devoted to the study of these small gap semiconducting
tubes, which has led to a good basic understanding.  Already 
from numerical calculations\cite{buck book} it has been known for a while that
only armchair tubes retain zero gap and therefore are truly metallic
while chiral tubes open small gaps because of the intrinsic curvature.
 These gaps depend on the radius and the chiral angle and
were found to be lower than room temperature for
tubes with radius exceeding 6\AA. Other numerical studies also examined the
effect of structural deformation such as stress\cite{heyd 97,yang 99} and
twist\cite{yang 99,rochefort 99} on the electronic band gap of 
nanotubes where it was found that the armchair tubes open a gap under
twist but are not affected by the uniaxial stress while an opposite
response arises in zigzag tubes. Analytic progress was
made by Kane et al\cite{kane 97} where 
the effects of deformations were modeled as a perturbing vector potential 
in an effective mass Hamiltonian.   Those calculations were
able to verify the numerical findings for the gaps from the intrinsic 
curvature as well as twists.  
After the completion of this work we also became aware
of a more recent analytic work by Yang et al\cite{yang 2000} where
both twist and stress on the nanotubes were considered (but not the
intrinsic curvature).

From the previous 
works \cite{buck book,heyd 97,yang 99,rochefort 99,kane 97,yang 2000} 
the gaps can in principle be calculated for almost any shape
of nanotube either by using a numerical method or by determining the 
exact metric and curvature tensors.   We now reconsider the effects of
deformations on primary metallic nanotubes and formulate the problem
in terms of a general symmetry breaking in the tight binding model.
This gives a straightforward analysis of the effect which results in 
a surprisingly simple and useful formula for the gap.  Our compact 
expression combines the effects from
intrinsic curvature, twist and stress and is
only a function of the chiral wrapping vector  $(n,m)$ of the tube.
This gives a quick and easy way to determine the gap and 
allows for a good insight in the physical effects as we will 
describe below.  Our results are a direct consequence of {\it any}
symmetry breaking effect in the tight binding model of the graphite
sheet in carbon nanotubes.

Our starting point is the observation that the energy separation between
the bonding and antibonding bands according to the graphite tight binding
scheme is
$2\gamma\sum_{i=1}^{3} e^{i\vec{k}\cdot\vec{R}_{i}}$,
 where $\vec{R}_{i}$ are the nearest neighbor bond vectors and
 $\gamma$ is the transfer integral which is the nearest neighbor 
Hamiltonian matrix element.\cite{wallace 47}
Since at zero temperature the bonding band is occupied
and the antibonding empty, the Fermi points $\vec K_F$ 
lie at the band crossings,
 which are, for the unperturbed graphite, the six corners of the
 hexagonal first Brillouin zone.   If we now break the symmetry of graphite
and allow different transfer integrals $\gamma_i$ depending on the direction
of the bonds $\vec R_i$ we arrive at a more general equation for the 
$k$-vectors at which the bands cross\cite{deltar}
\begin{equation}\label{Eq. 1}
\sum_{i=1}^{3}\gamma_{i} e^{i\vec{k}\cdot\vec{R}_{i}}=0,
\end{equation}
which defines the points of zero gap in $k$-space $\vec K_F'$.
For small changes $\gamma_{i}=\gamma +\delta\gamma_{i}$ we expect small 
shifts in the band crossing location $\vec{K}_{F}'=\vec{K}_{F}+\Delta\vec{k}$.
Since we are dealing with primary metallic tubes, we know 
that $\sum_{i=1}^{3} e^{i\vec{K}_{F}\cdot\vec{R}_{i}}=0 $
 where $\vec{K}_{F}$ are the unperturbed Fermi points. 
Working in the nanotube coordinates $(\hat{c},\hat{t})$ adopted from  
 Ref.~\onlinecite{yang 99} where
$\hat{c}$~is the circumferential direction and $\hat{t}$ stands for
the translational direction along the axis, the bond vectors
are 
\begin{equation}
\begin{array}{l}
\vec{R}_{1}=\frac{a}{2 c_{h}} [ (n+m)\hat{c} - \frac{1}{\surd{3}}(n-m)\hat{t} ]\\
\vec{R}_{2}=\frac{a}{2 c_{h}} [ -m\hat{c} + \frac{1}{\surd{3}}(2n+m)\hat{t} ]\\
\vec{R}_{3}=\frac{a}{2 c_{h}} [ -n\hat{c} - \frac{1}{\surd{3}}(n+2m)\hat{t} ],
\end{array}
\end{equation}
where $a \approx 2.49\rm \AA$ is the length of the honeycomb unit vector and
$c_{h}=\sqrt{n^{2}+nm+m^{2}}$ is the circumference in units of $a$.
Both inequivalent Fermi points in graphite 
give the same result when estimating the gap
so it is sufficient to consider just one of them.  For an unperturbed 
Fermi point we write
\begin{equation} \label{kf}
\vec{K}_{F}=\frac{2\pi}{3ac_{h}}[(m+2n)\hat{k}_{c}+m\sqrt{3} \hat{k}_{t}],
\end{equation}
where $\hat{k}_{c}$ and $\hat{k}_{t}$ correspond to the 
$k$-vectors along the 
circumferential and translational directions, respectively. 

In order to get the gap, we now need to know the
distance between the new Fermi point to the nearest quantization
line at the quantized circumferential $k_c$-values. 
Since the quantization lines are parallel to $\hat{k}_t$, this 
distance is given by $\Delta k_{c}$ the shift along the circumferential
 direction $\hat{k}_{c}$.
Expanding  Eq. (\ref{Eq. 1}) to linear order in the 
perturbations $\delta \gamma_i \equiv \gamma_i - \gamma$ we find
\begin{equation}\label{deltakc}
\Delta k_{c}=\frac{1}{ac_{h}\gamma\sqrt 3}[\delta
\gamma_{1}(m-n)+\delta \gamma_{2}(2n+m)-\delta \gamma_{3}(n+2m)]
\end{equation}
The gap is then obtained by exploiting the fact that close to the 
Fermi point the dispersion relation is linear and 
isotropic\cite{wallace 47}
\begin{equation}\label{gap}
E_g = \sqrt{3} a \gamma |\Delta k_{c}|. 
\end{equation}

We now want to determine the changes to the transfer integrals
$\delta\gamma_{i}$ due to the curvature and deformations. To first order
this change is 
proportional to the change of the bond length between two neighboring
carbon atoms, but may also be created by a misalignment of two neighboring 
$\pi$ orbitals.  In general we find that we can always express the change
of the nearest neighbor transfer integrals in terms of a bond
deviation matrix \textbf{D}
\begin{equation}\label{delta gamma}
\delta \gamma_{i}=\vec{R}_{i}\cdot\textbf{D}\cdot \vec{R}_{i}/R_i^2.
\end{equation}
This deviation matrix is useful for describing the effects of 
stress, twists, and curvature in a simple unified way as we will 
see below.
All nanotubes have an 
intrinsic curvature which causes hybridization of the otherwise
orthogonal $\pi$ and $\sigma$ orbitals. Since the $\sigma$ bands lie
normally far from the Fermi energy, we only consider the
hybridization effect 
on the $\pi$ band, which crosses the Fermi point in primary metallic tubes
as long as the tube's radius is $R \agt 2.4 \rm \AA$.\cite{comment}

Following the calculations of Slater and Koster\cite{slater koster}
we can assume that the transfer integrals are proportional to
cosine of the misalignment angle $\phi$ between two neighboring
$\pi$ orbitals.  (A calculation that takes into account the
full rehybridization of all orbitals will be considered in a future
study.)
 Using $\cos \phi \approx 
1-\frac{1}{8}\left(\frac{a_{c-c}}{\rm{R}}\right)^{2}$, we can immediately 
express the deformation tensor for the intrinsic curvature in the basis
of $\hat c$ and $\hat t$ 
\begin{equation} \label{Dcurv}
\textbf{D}^{\rm curv}= \left [\begin{array}{cc}\frac{\gamma}{8}\left(\frac{a_{c-c}}{\rm{R}}\right)^2 & 0
    \\ 0 & 0 \end{array}\right ],
\end{equation}
where $a_{c-c}=a/\sqrt{3}$ is the carbon-carbon bond length and
${\rm R}=ac_{h}/2\pi$ is the tube's radius. From 
Eqs.~(\ref{deltakc}-\ref{Dcurv}) we find
\begin{equation}\label{curvature} 
E_g=\frac{\gamma \pi^{2}}{8 c_{h}^{5}} (n-m)(2n^{2}+5nm+2m^{2}).
\end{equation}
This formula agrees remarkably well with previous numerical 
studies\cite{buck book} if we choose $\gamma = 2.5$eV and
also agrees with the results of Ref.~\onlinecite{kane 97}.
Equation (\ref{curvature}) gives the energy gap of all primary metallic
nanotubes without any applied deformations.
One observes that the
armchair nanotubes $m=n$ are the only real metallic tubes,
while the primary metallic zigzag tubes ($m=0$),~open the highest gaps.


Next we want to examine the effect of a general two dimensional
structural deformation such as twists and stress 
on the gap in the primary metallic tubes.
Our deformation can be written as
$\vec{R}\rightarrow(\textbf{I}+\textbf{S})\vec{R}$, where $\vec{R}$ 
is any vector on the tube's surface, $\textbf{I}$
is the identity matrix and $\textbf{S}$ is the deformation matrix
\begin{equation}
\textbf{S}\equiv\,\left[\begin{array}{cc}\ \epsilon_{c}\  & \ \xi\ 
    \\ 0 & \epsilon_{t} \end{array}\right].
\end{equation}
Here $\epsilon_{c}$ and $\epsilon_{t}$ are uniaxial stresses along
the circumferential and the translational directions and $\xi$ is the
strain (nanotube twist).
The bond deviation matrix is then given by 
$\textbf{D}^{\rm deform}=|\vec{R}|\,b\,\textbf{S}$, where
$b\simeq 3.5$\,eV/\AA~is the
linear change in the transfer integral with a change in the bond
length $\gamma_{i}\rightarrow\gamma_{i}+b\,|\Delta\vec{R}_{i}|$, and
$|\vec{R}|=a/\sqrt{3}$ is the bond length.  
 We now use $\textbf{D}^{\rm deform}$ to
 obtain the $\delta\gamma_{i}$ of Eq.~(\ref{delta gamma}) and as we did
 with the curvature, inserting in
 Eq.~(\ref{deltakc}) and using the dispersion relation we find
\begin{eqnarray}
E_{g}& =& \frac{a b}{4c_{h}^{3}}\,\left|\,\sqrt{3}\,
(n-m)(2n^{2}+5nm+2m^{2})(\epsilon_{c}-\epsilon_{t}) \right.\nonumber\\
&~&- \left.9nm(n+m)\xi\,\right|.
\label{deformation}
\end{eqnarray} 
This equation is the response to a two-dimensional linear deformation within  
the graphite sheet. We notice from
Eq.~(\ref{deformation}) that in
the presence of equal uniaxial stresses in both directions $\hat{c}$ and
$\hat{t}$, a gap is not opened, as expected since the bonds would
maintain their symmetry. The response of armchair and zigzag tubes 
is complimentary as noticed previously in the numerical
studies,\cite{yang 99}
i.e. zigzag tubes have the maximum sensitivity  for a uniaxial
deformation ($\epsilon_{c}$ or $\epsilon_{t}$) but are insensitive to twists 
$\xi$, while the opposite is true for armchairs, reaffirming that
a twist deformation is the only possible source for an energy gap in the
armchair tube.

A uniaxial stress $\epsilon_{c}$ around the
circumference corresponds to a global change of radius. This effect 
can come about as a time dependent deformation due to lattice
vibrations in the breathing mode. 
Realistic static deformations on the other hand correspond to the 
intrinsic curvature, a stress along the tube $\epsilon_t$ and  a twist $\xi$.
Therefore the total gap from static deformations is given by combining 
Eq.~(\ref{curvature}) and Eq.~(\ref{deformation}) with
$\epsilon_c=0$.
 The total band equation now reads
\begin{eqnarray}
E_{g}& =& \left|\left(
 \frac{\gamma \pi^{2}}{8 c_{h}^{5}} - 
\frac{a b \sqrt{3}}{4 c_{h}^{3}}\epsilon_{t} \right)
(n-m)(2n^{2}+5nm+2m^{2}) \right.\nonumber\\ &~&- \left. 
 \frac{9 a b}{4 c_{h}^{3}}nm(n+m)\xi\,\right|
\label{total gap nm}
\end{eqnarray} 
which is the main result of our paper.

In some cases it may be useful to express this formula in terms 
of the chiral angle $\alpha$ and the radius $\rm{R}$ of the tube,
which gives
\begin{equation}\label{total gap}
E_g = \left|\left(\frac{\gamma a^{2}}{16 {\rm R}^{2}} -
\frac{ab\sqrt{3}}{2}\,\epsilon_{t}\right)\,\sin{3\alpha} 
  -\,\frac{ab\sqrt{3}}{2}\,\xi\,\cos{3\alpha}\,\right|.
\end{equation}
In this form our results are then consistent with previous
calculations\cite{kane 97} where 
twists and the intrinsic curvature (but not stress) have been considered
as a function of the chiral angle. After the completion of this work,
 a paper deriving the change in the
band gap due to deformations was published,\cite{yang 2000} which is also
consistent with the angular dependence of the
deformation part  of Eq.~(\ref{total gap}) (i.e. without the
intrinsic curvature).

\begin{figure}
\epsfxsize=3.35in
\epsfbox{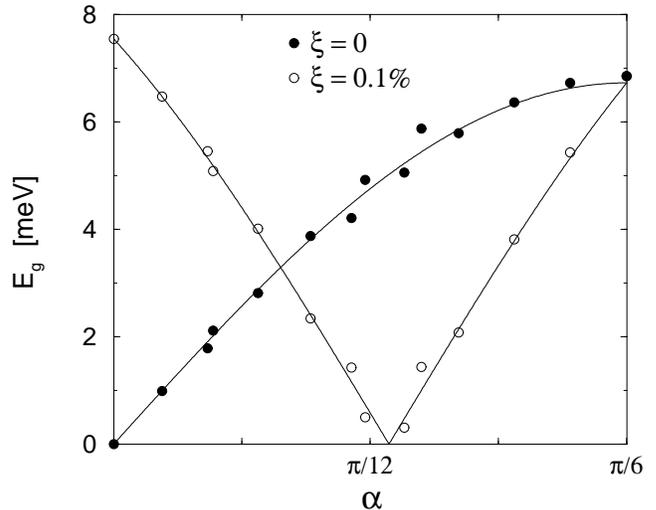}
      \caption{\label{two_rad}Energy gaps of primary metallic tubes of radius
        $12\pm 0.5$\AA~with no twist $\xi = 0$ and a twist $\xi = 0.1\%$ as a
        function of chirality. The continues lines correspond to the
        average radius ($12$ \AA) while the circles
        correspond to the actual tubes. A very small twist can thus
        have a profound effect on the energy gaps.}
\end{figure}

We see that Eq.~(\ref{total gap}) shows an interesting interplay 
between curvature and deformation effects as plotted in 
in Fig.~\ref{two_rad}. A very small twist can 
actually remove the gap due to the intrinsic curvature.  For a given
radius the gapless state is therefore moved  
towards tubes with higher chirality.

In summary, we have presented a straightforward procedure to calculate the
energy gap induced by a general broken bond symmetry.
This leads to a simple analytic expression for the 
 band gaps from both the intrinsic curvature and applied deformations, 
which provides a quick and reliable way 
to estimate the physical effects.   
 These gaps  have important consequences since 
they are generally of the same order as room temperature for most
 primary metallic SWNT.   Only armchair tubes are generically gapless, but 
a very small twist induces a gap of the order of other small gap semiconducting
nanotubes.  Such a small twist, on the other hand, moves the
gapless state to tubes with higher chirality.

\begin{acknowledgements}
We would like to thank Vitali Shumeiko for inspirational discussions.
This research was supported in part by the Swedish Natural 
Science Research Council.
\end{acknowledgements}



\begin{thebibliography}{99}
\bibitem{deltar}
In general we will also consider different bond vectors $\vec R_i'$. 
However, this simply corresponds to a change of the coordinate system in
both real and reciprocal spaces with a conserved dot product 
$\vec k\cdot \vec R$.  The distance to the quantization lines in 
Eq.~(\protect{\ref{deltakc}}) is therefore not affected by different bond
vectors, but only by the change in the transfer integrals.
\end{thebibliography}
\end{document}